%% file: main.tex
\documentclass[sigconf]{acmart}
\usepackage{booktabs} 
\usepackage{xspace}
\usepackage{subfig}
\usepackage{multicol}
\usepackage{wasysym}
\usepackage{bmpsize}

\usepackage{subfig}
\setcopyright{rightsretained}

\acmDOI{10.475/123_4}

\acmISBN{123-4567-24-567/08/06}

\acmConference[RecNLP 2019]{AAAI 2019}{January 27-28, 2019}{Hawaii, USA}
\acmYear{2019}
\copyrightyear{2019}

\acmArticle{4}
\acmPrice{15.00}

\usepackage{graphicx}
\usepackage{amsfonts} 
\usepackage{xcolor}

\usepackage{CJKutf8}
\usepackage{multirow}
\usepackage{multicol}
\usepackage{xspace}
\usepackage{subfig}
\usepackage{amsmath}

\newcommand{\system}{TextCNN\xspace}
\newcommand{\phsystem}{PhrecSys\xspace}   
\newcommand{\baseline}{RNN-CCE\xspace}

\usepackage{url}
\begin{document}
\title{Enriching Article Recommendation with Phrase Awareness}

 \author{Chia-Wei Chen}
 \affiliation{
   \institution{Academia Sinica}
   \city{Taipei}
   \state{Taiwan}
 }
 \email{ss87021456@iis.sinica.edu.tw}

 \author{Sheng-Chuan Chou}
 \affiliation{
   \institution{Academia Sinica}
   \city{Taipei}
   \state{Taiwan}
 }
 \email{angelocsc@iis.sinica.edu.tw}

 \author{Lun-Wei Ku}
 \affiliation{
   \institution{Academia Sinica}
   \city{Taipei}
   \state{Taiwan}
 }
 \email{lwku@iis.sinica.edu.tw}

\begin{abstract}
Recent deep learning methods for recommendation systems are highly sophisticated.
For article recommendation task, a neural network encoder which generates
a latent representation of the article content would prove useful. However,
using raw text with embedding for models could degrade sentence meanings
and deteriorate performance. In this paper, we propose \phsystem (Phrase-based Recommendation System), which injects phrase-level features
into content-based recommendation systems to enhance feature informativeness
and model interpretability. Experiments conducted on six months of real-world
data demonstrate that phrase features boost content-based models in predicting
both user click and view behavior. Furthermore, the attention mechanism
illustrates that phrase awareness benefits the learning of textual
focus by putting the model's attention on meaningful text spans, which leads to interpretable
article recommendation.

\end{abstract}



\settopmatter{printacmref=false}
\maketitle
\begin{CJK*}{UTF8}{bsmi}

\section{Introduction}

Learning from humans has shown to be effective for designing better AI
systems~\cite{guu2017generating}.
Intuitively, humans cannot memorize lengthy article information, which
serves as a clue for article recommendation systems. That is, after reading
through an article, there is a high possibility that only a few parts of the
article will attract user attention. However, deep learning models that adopt
raw text with embedding might degrade the meaning of sentences, which violates this precept. 

The phrase, a small group of words standing together as a conceptual unit, can
serve as a useful component of these attractive parts, or chunks. Strong data-driven 
methods for extracting quality phrases have been developed 
recently~\cite{autophrase, segphrase}. With these, high-quality phrases can be extracted
without additional human effort. However, despite the regular utilization of handcrafted
keywords, article recommendation seldom leverages phrase mining.
Therefore, we propose \phsystem, which utilizes phrase-level
features in content-based recommendation systems. 
We add phrase mining to state-of-the-art content-based
recommendation models and compare their performance with the original
models. Moreover, we use the attention
mechanism~\cite{structured-self-attention} and visualize the changes in
attention weights during training for a clear view of the model focus.

This work includes two contributions. First, to the best of our knowledge, we
are the first to incorporate phrase mining techniques to boost content-based
article recommendation systems. Second, we visualize the change of attention
weights during training to show systems when given phrase features: systems
learn to read in a more focused way. As a great output, these focused chunks
also help us interpret how these systems are producing recommendations.

\section{Related Work}

As the recommended targets are news articles, and the system is attempting to
recommend articles based on the article currently being read, content-based recommendation models
are the most related. Content-based models inherently sidestep both the
cold-start issue and the problem of proposing unrelated content, as the content
is the primary information needed to make good recommendations. This is
especially useful for article recommendation in environments where new articles
arrive regularly. 
Conventionally, articles are treated as bags of words, whose similarity is
calculated by scoring functions such as TF-IDF~\cite{pazzani2007content} and
BM25~\cite{BM25}. Recent advances in textual feature extraction with NN
models~\cite{conneau2017very, le2014distributed} make it possible to generate
article representations using NN encoders~\cite{chen2017joint}. With the help
of word embedding, it is now possible to aggregate various neural network
architectures in the textural features for different tasks. Information
retrieval models can also be used for recommendation applications~\cite{cdssm,
knrm, mvlstm}, additionally incorporating user feedback~\cite{ni2017estimating}
and past behavior~\cite{debnath2008feature} to generate personalized
recommendations.

We use phrase mining as a technique to extract the concepts that stick in readers'
minds after reading. Phrase-level features have proved useful in
many different natural language tasks. 
\citeauthor{LAKI}~\cite{LAKI} use phrases to enhance the interpretability of
domain-specific documents, and demonstrate effectiveness and efficiency via
various experiments.
\citeauthor{phrase-topic_model}~\cite{phrase-topic_model} proposes a phrase-level topic model 
which improves semantic coherence within each topic. Further, he conducts a human
study which shows that the outputs of the phrase-level topic model are easier to
understand. 
\citeauthor{phrase-translate}~\cite{phrase-translate} propose a method to translate phrases in neural
machine translation (NMT), enhancing the BLEU scores on a translation task from
Chinese to English. In this paper, we leverage the power of phrases in briefing
and interpreting to build a better article recommendation system.

\section{System Overview}

\begin{figure}[!hbt]
\vspace{-1pc}
\includegraphics[scale=0.4]{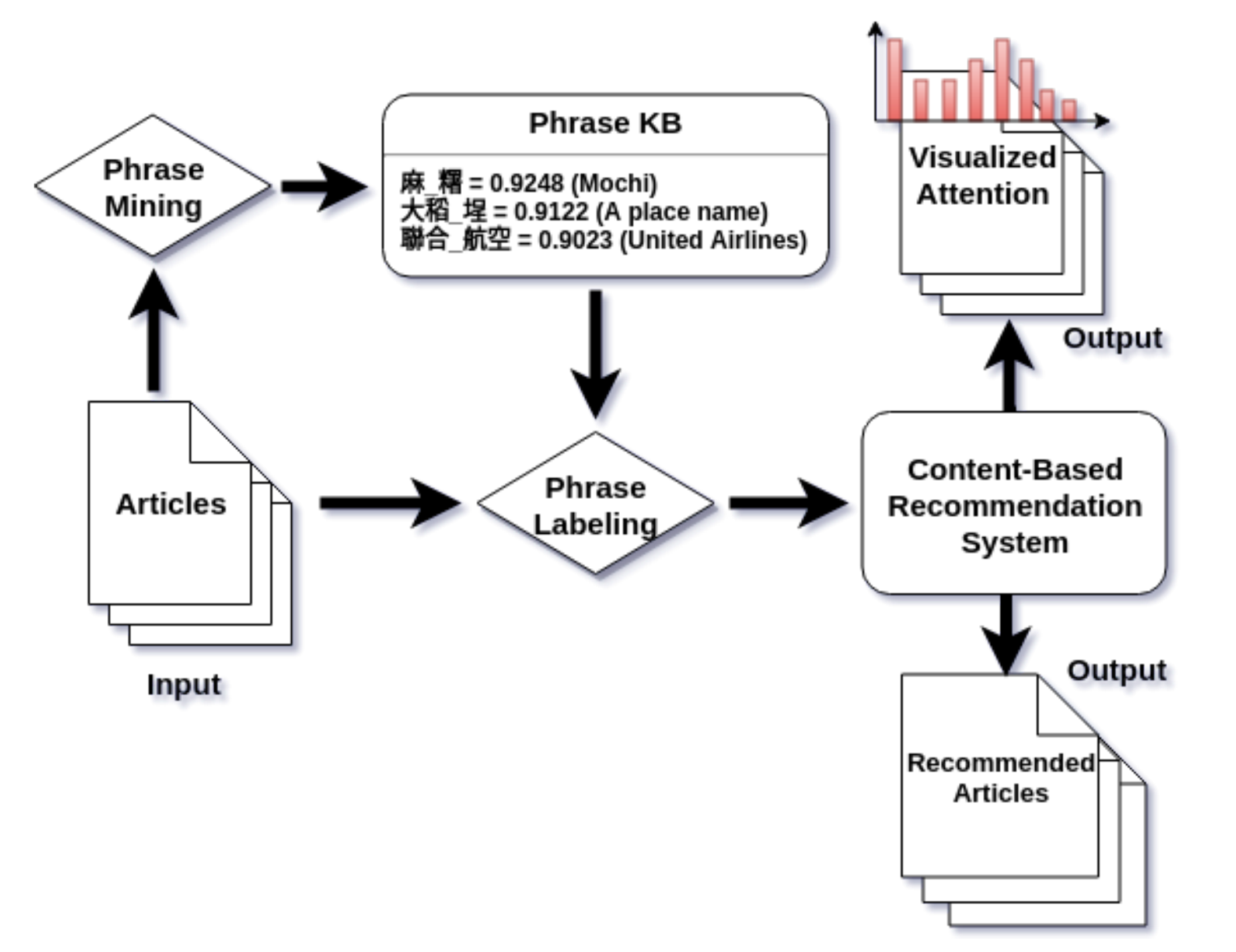}
\caption{\phsystem system overview}
\label{fig:System_overview}
\vspace{-1pc}
\end{figure}

An overview of our \phsystem is shown in Fig.~\ref{fig:System_overview} The first
step is extracting high-quality phrases in each article. For the
efficiency and scalability needed by an automatic general article
recommendation system, we use Autophrase~\cite{autophrase}, a data-driven and
distant supervised learning algorithm. 

Phrases from Autophrase with scores no lower than 0.5 are collected as
candidates for phrase labeling. In this step, the phrase with the longest
length is given a higher priority, i.e., longest match. Finally, phrases are
labeled in each article.

The labeling process yields phrase-based articles, that is, articles injected with 
phrase information. We apply Glove~\cite{pennington2014glove}
to train the phrase embeddings for the next step. Note that punctuation is dropped in articles.

The five content-based recommendation models listed below are utilized to evaluate
whether the added phrase mining component really benefits recommendation
performance: \system, CDSSM, MV-LSTM, K-NRM, and BiLSTM-SA. Given the article currently
being read and a pool of candidate articles, these models generate
the recommendation list ordered by recommendation score:
\label{Baseline}

\textbf{\system}. Similar to~\cite{kim2014convolutional}, convolutional neural 
networks with different kernel sizes are used to mimic unigram, bigram, and trigram
language models, followed by a max-pooling layer and an asymmetric cosine
similarity layer to calculate the score between two articles. \\
\indent \textbf{CDSSM}~\cite{cdssm}. Uses a single convolutional neural network to
learn low-dimensional semantic vectors followed by multi-layer perceptrons
(MLP), after which the normalized dot product is used to calculate the score between two articles. \\
\indent \textbf{MV-LSTM}~\cite{mvlstm}. Incorporates Bi-LSTM to capture contextualized
local information in each positional sentence and then uses the normalized dot
product to generate an interaction tensor. K-Max pooling is used to obtain the
top-$k$ important information, after which MLP is used to produce the final
score.  \\
\indent \textbf{K-NRM}~\cite{knrm}. A kernel-based neural model for document ranking
which uses a translation matrix that models word-level similarities via word
embeddings and utilizes kernels to extract multi-level soft-match features. A ranking layer is followed to generate ranking scores. \\
\indent \textbf{BiLSTM-SA}~\cite{structured-self-attention}. Uses Bi-LSTM followed by a
self-attention mechanism comprised of two linear layers with a softmax
layer to generate multiple probabilities for each timestep.

In addition to the recommendation task itself, \phsystem can visualizes the learned
focus of each article via the attention technique, which helps models learn to
focus on important parts while dealing with long sequences such as news
articles. We then observe how the model learns with features of different
granularity. Here we select BiLSTM-SA~\cite{structured-self-attention} 
for attention visualization: we store the attention weights learned in
each training epoch, and then plot the tokens in the current article with
different color depths according to their attention weights. Given these different 
color depths, we can easily perceive which part of the article the model is
focusing on for recommendation in the training process.
Figure~\ref{attention_visualization} shows an example; more details are provided below.

\section{Dataset}
\label{Data Preparation}

Two datasets~-- Central News Agency (CNA)\footnote{http://www.cna.com.tw} and Taiwan People News (TPN)\footnote{http://www.peoplenews.tw} -- were provided by our collaborating partner cacaFly\footnote{https://cacafly.com} for experiments. 
TPN and CNA are Taiwanese news websites articles written in Mandarin.

Two types of user behavior~-- clicks and views~-- are logged.
Specifically, a \emph{View} log is recorded when the user reads the current article
$a_c$. When the user scrolls to the bottom of the web page to reach the
recommendations, the partner system interprets this to mean that the user has
finished reading $a_c$; a \emph{Click} log is recorded when the user clicks on any
of the recommendations. We generate \emph{Click} log sequences to predict the
articles that the user will click on next, and \emph{View} log sequences to predict
the articles that will be read next, accordingly.

\input{tables/0_data_stat.tex}

For a \emph{Click} log sequence, a positive article pair ($a_c$, $\hat{a_r}$) is
generated from the article sequence if $a_i$ appears in the \emph{View} log and $a_{i+1}$
in the \emph{Click} log, $1 \leq i \leq n-1$. 
As recency is often taken into account for
recommendation~\cite{ahmed2012fair,doychev2014analysis}, we generate no more
than the latest 8 positive pairs per user for experiments if the user has read
many articles. 
For the \emph{View} log sequence, we select those articles in the article sequence
with the \emph{View} log, and pair them to generate it. That is, ($a_c$,
$\hat{a_r}$) is a pair of successive articles that are both being read by the
user. Along with each positive pair, there are $m$ negative pairs ($a_c$,
$\check{a_r}$), where the $\check{a_r}$'s are the un-clicked recommended
articles for $a_c$.

Let $l$ be the number of article pairs in each user's log sequence.
For each user's \emph{Click} or \emph{View} log sequence, the last article pair ${(a_c,
a_r)}_l$ is for testing, the second-to-last is for validation, and the rest are
for training. Table~\ref{tb:data_stat} shows the statistics of the training,
validation, and testing data.

\section{Experiments and Results}
\label{sec:exp}

In this section, we first describe the datasets, model settings, and evaluation
metrics, after which we discuss the performance and present the visualization of
attention weights during training on word-level and phrase-level systems.

\input{tables/8_nce_data_stastic.tex}

\subsection{Settings and Metrics}
\label{sec:exp_setting}
For each dataset, we pre-train 50-dimensional phrase embeddings and
word embeddings with GloVe~\cite{pennington2014glove} separately, using only
the articles within each dataset.
The total number of articles used to train word/phrase embeddings for each
dataset are listed in Table~\ref{tb:nb_article}. Below we describe the parameter settings
for different recommendation systems. Note that hinge loss is used as objective function. \\
\indent\textbf{\system}. We set CNN filter number to 32, the maximum article length to 512, and $\alpha$ in the asymmetric cosine to 0.85.
During training we used the Adam optimizer with a learning rate of 0.1. \\

\indent\textbf{CDSSM}. We set the CNN filter number to 32 and selected 128 as the MLP output
dimension for article semantic representation. During training we used the Adam optimizer with a learning
rate of 0.01. \\
\indent\textbf{MV-LSTM}. We set the LSTM hidden size to 32 and the top-$k$ value to 512. Two
layer MLPs were applied at the end of model, and 64 and 1 were the number of output
dimensions respectively. During training we used the Adam optimizer with a learning
rate of 0.01. \\
\indent\textbf{KNRM}. We set the kernel number to 32 and the sigma value to 0.05. 
During training we used the Adam optimizer with a learning rate of 0.001. \\
\indent\textbf{BiLSTM-SA}. We set the LSTM hidden size to 32 for each direction,
$d_a$ to 100 for the fully-connected layer inside self-attention, and the attention number
$r$ to 15. During training we used the Adam optimizer with a learning rate of
0.01.

We adopted several evaluation metrics to fairly evaluate the recommendation performance.
To evaluate the performance of recommendation, per the literature we used the
mean reciprocal rank (MRR)~\cite{voorhees1999trec}, 
accuracy (Acc)~\cite{devooght2017long}, hit ratio at 3 (h@3), and hit ratio at
5 (h@5)~\cite{kumar2017deep}. The hit ratio h@k intuitively measures whether the test item is
present in the top-$k$ list. 
\begin{equation}
\begin{split}
    Acc &= \frac{1}{|U|}\sum_{u = 1}^{|U|} accuracy(u) \\
    accuracy(u) &=
    \begin{cases}
        1, & \text{if}\ Rank(score_{c,\hat{r}}) = 1 \\
        0, & \text{otherwise}
    \end{cases}
\end{split}
\end{equation}
\begin{equation}
\begin{split}
    h@N = \frac{1}{|U|}\sum_{u = 1}^{|U|} h(u, N). \  
    h(u, N) =
    \begin{cases}
        1, &\text{if}\ Rank(score_{c,\hat{r}}) \leq N \\
        0, & \text{otherwise}
    \end{cases}
\end{split}
\end{equation}
Assuming $\left|U\right|$ testing instances in the testing set, each
testing instance $u \in U$ contained one positive pair ($a_c$, $\hat{a_r}$) and
$m$ negative pairs ($a_c$, $\check{a_r}$).
The average MRR was calculated as
\begin{equation}
    MRR=\frac{1}{|U|}\sum_{u = 1}^{|U|}\frac{1}{Rank(score_{c,\hat{r}})},
\end{equation}
where $Rank(score_{c,\hat{r}})$ was the rank of the positive pair score among
all scores of pairs in $u$.
\vspace{-1em}

\input{tables/4_click_vs_view.tex}  
\begin{figure*}[!tbp]
  \vspace{-1.5em}
  \centering
  \subfloat[Phrase]{\raisebox{-0ex}{ \includegraphics[height=0.25\textheight,width=0.49\textwidth]{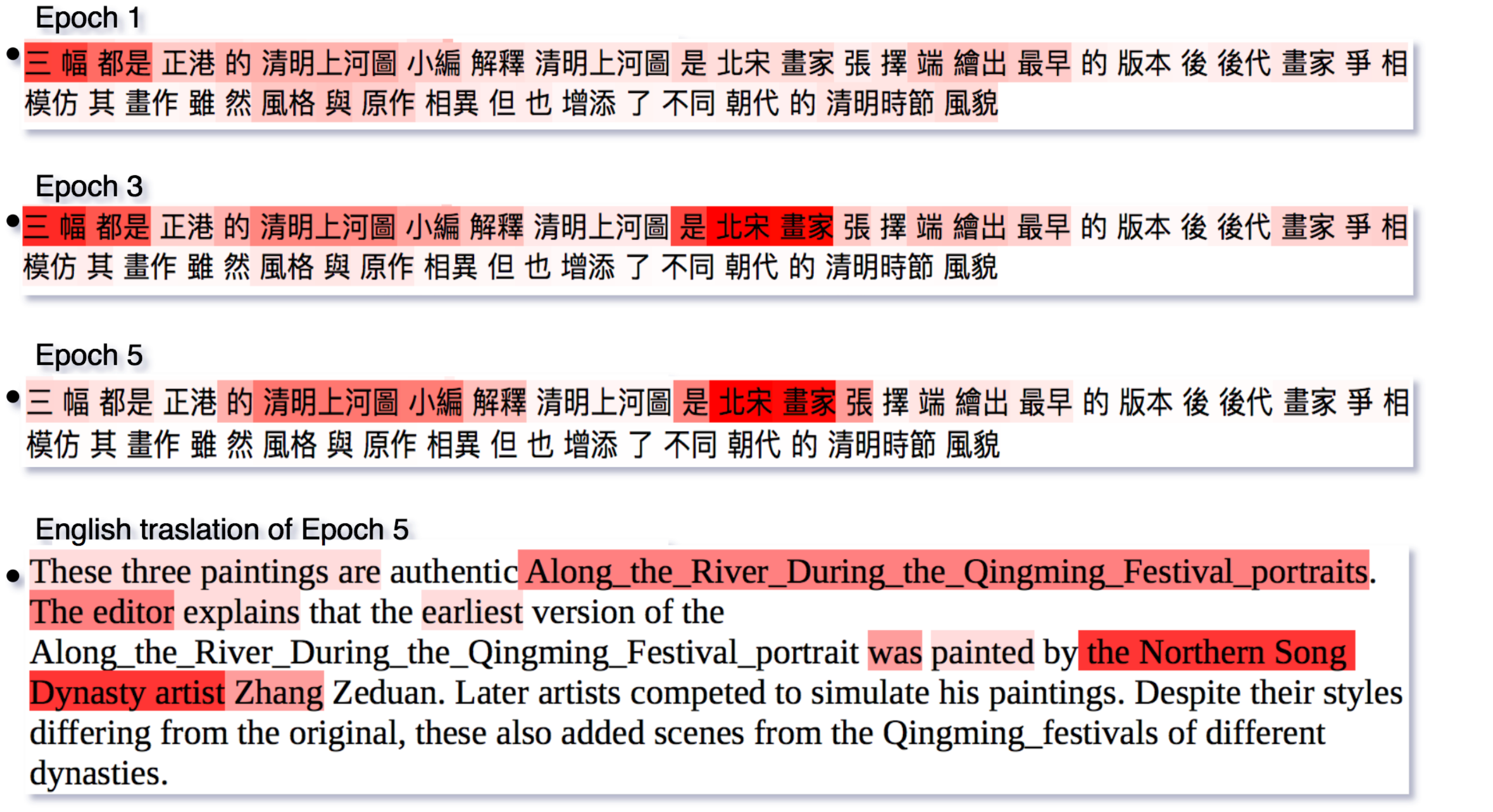}}\label{fig:f1}}
  \hfill
  \subfloat[Word]{\includegraphics[height=0.25\textheight,width=0.49\textwidth]{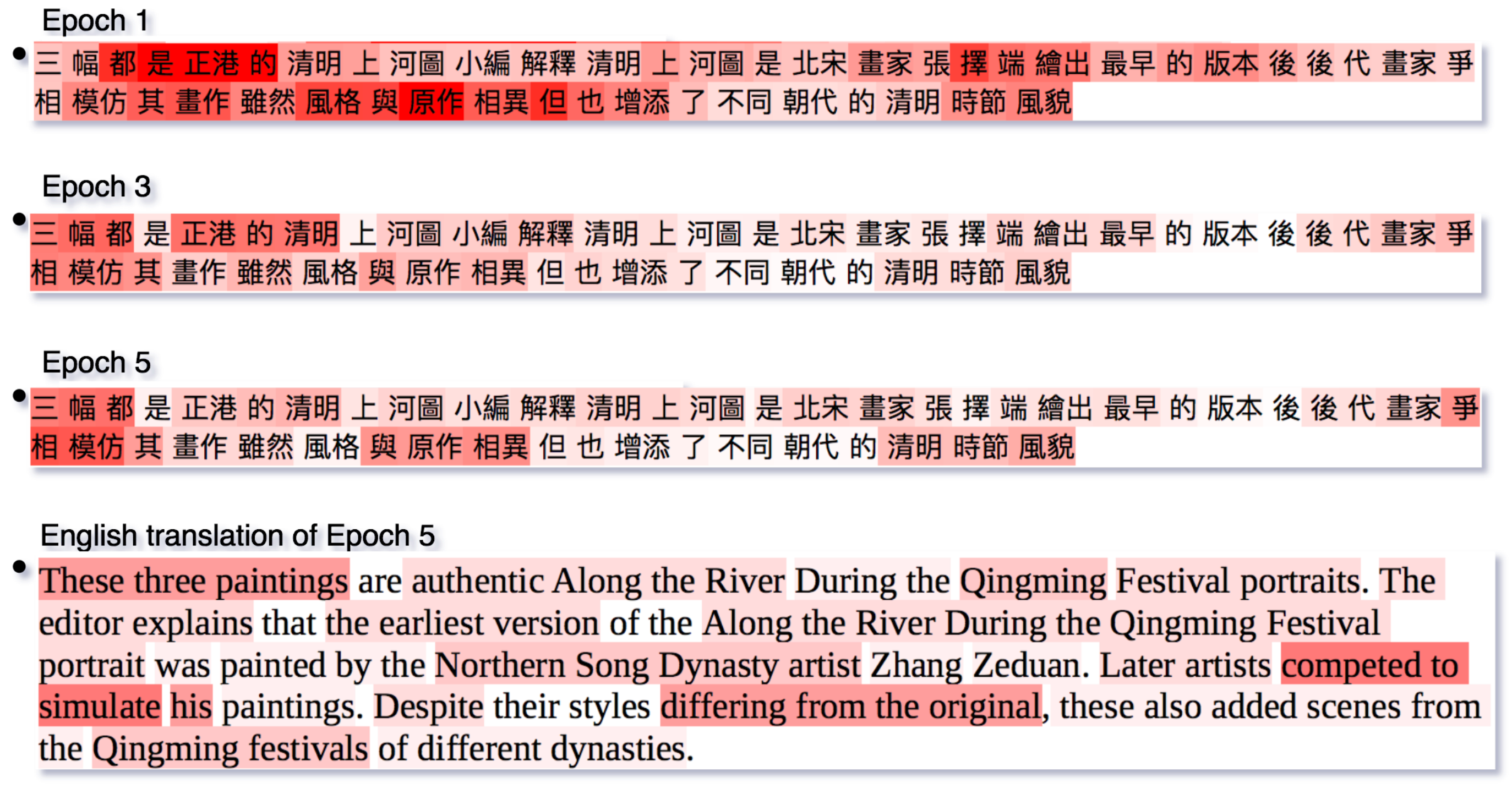}\label{fig:f2}}
  \caption{Changes in attention weights during training process}
  \label{attention_visualization}
  \vspace{-1.6em}
\end{figure*}


\subsection{Recommendation Performance}

We examined the performance on two datasets (\emph{News\_1}, \emph{News\_2}), both of which
exhibit two types of
behavior~-- \emph{Click} and \emph{View}. The five models described in
Section~\ref{Baseline} and ~\ref{sec:exp_setting}  were evaluated to compare their performance with and
without phrase mining: the results are shown in Table~\ref{tb:click}.

For the prediction of the next clicked article (the article that the user
clicks on next), most models are slightly enhanced by phrase mining. 
However, for the next viewed article (the article that the user reads next),
the performance of all models is improved significantly when leveraging
phrase-level features.
This may be because text content might not be the first
priority to users when clicking, as eye-catching titles or images may hold a
special attraction. In this case, though phrases capture the key points
of an article, their benefit is limited.
For view prediction, in contrast, the text content is essential. We observe
that a large amount of views come from websites other than the current news
media and very often from search engines. In these cases, the users' intention
is clearly a specific topic that interests them and hence they tend to read
contents that are logically related successively. Another reason is that as
users read through the current article, its content becomes a important factor
in the surfing history. The notable enhancement of phrase-level features
shows that phrases can enhance the model's ability to learn the relatedness of
contents between articles.

\subsection{Visualization}
To visualize the focus points being learned by the model, 
we record the attention weights of composite units (words or phrases plus
words) of the current article in each training epoch and observe their
change over time. Figure~\ref{attention_visualization} shows the effect of phrase
information. Figure~\ref{fig:f1} is the article parsed and trained with
phrase-level units, whereas Figure~\ref{fig:f2} is the article parsed and trained
with word-level units. The attention weights of epoch 5 illustrate the overall
result: when phrase information is provided, attention weights
saturate to a reasonable number of units. In comparison, without phrase mining, 
weights are evenly distributed, with no obvious key units to be found in the 
visualization. 

Comparing Figure~\ref{fig:f1} and Figure~\ref{fig:f2} from Epochs 1 and 3 to Epoch
5, we observe that the attention weights are both evenly distributed and show no
particular focus on anything in the beginning. Later, in Epoch 3, training
with phrase-level features causes weights to be more concentrated and accurately focused
on meaningful parts; finally, in Epoch 5, training with phrase-level
information successfully leads to fewer but more informative cues. We conclude
that phrases contain more compact information and are more easily utilized for
recommendation models.

\section{Conclusion}
In this paper, we use data-driven phrase mining to
automatically extract high-quality phrases for news article recommendation. We
show that the proposed approach yields improvements in both click and view prediction.
Additionally, we visualize the learning process of both phrase-level and
word-level models to illustrate that the merit of phrase mining for
recommendation is its ability to put the focus on key units. With this advantage,
the visualization itself can be interpreted and is of great value to
public opinion analytics.

\begin{acks}
  The authors would like to thank our collaborating partner cacaFly, the leader of digital advertising in Taiwan, for providing user website log in this paper. Research of this paper was partially supported by cacaFly, under the collaborative project contract O5T-1060629-1Cb.
\end{acks}

\end{CJK*}

\bibliographystyle{ACM-Reference-Format}
\bibliography{bibliography}

\end{document}

%% file: tables/0_data_stat.tex
\begin{table}[!hbt]
\vspace{-1em}
\small
\begin{tabular}{c|rrr|rrr}
    \multirow{2}{*}{Dataset} &  \multicolumn{3}{c|}{\emph{Click}} & \multicolumn{3}{c}{\emph{View}} \\ 
    \cline{2-7}
     & \multicolumn{1}{c}{Train} & \multicolumn{1}{c}{Val} & \multicolumn{1}{c|}{Test} & \multicolumn{1}{c}{Train} & \multicolumn{1}{c}{Val} & \multicolumn{1}{c}{Test} \\ 
    \hline
    CNA  & 28,950 & 4,825
    & 4,825 & 635,495 & 97,832 & 89,498 \\
    TPN & 17,041 & 2,841
    & 2,841 & 137,029 & 21,305 & 20,122
    \end{tabular}
\caption{\emph{Click} and \emph{View} article pairs}
\label{tb:data_stat}
\vspace{-2.5em}
\end{table}

%% file: tables/8_nce_data_stastic.tex
\begin{table}[h]
    \vspace{-1em}
    \begin{tabular}{c|c|c}
        Dataset  & CNA & TPN   \\
        \hline
        Articles & 257,705 & 54,033  \\
    \end{tabular}
    \caption{Number of articles in each dataset}
    \label{tb:nb_article}
\vspace{-3em}
\end{table}

%% file: tables/4_click_vs_view.tex
\begin{table*}[t]
    \small
    \centering
    \begin{tabular}{c|cccc|cccc|cccc|cccc}
        \multirow{2}{*}{Model}  & \multicolumn{4}{c|}{CNA-click} & \multicolumn{4}{c|}{TPN-click} & \multicolumn{4}{c|}{CNA-view} &
        \multicolumn{4}{c}{TPN-view} \\
        \cline{2-17}
         & MRR & Acc & h@3 & h@5 & MRR & Acc & h@3 & h@5 & MRR & Acc & h@3 & h@5 & MRR & Acc & h@3 & h@5 \\ 
        \hline
        K-NRM & \textbf{.340} & .132 & .393 & .612  & .282 & .095 & .281 & .477 & .263 & .087 & .259 & .428 & .293 & .104 & .309 & .495 \\
        MV-LSTM & .334 & .127 & .385 & .606 & .335 & .140 & .363 & .571 & .476 & .283 & .527 & .738 & .484 & .293 & .574 & .741 \\
        CDSSM & \textbf{.384} & \textbf{.169} & \textbf{.460} & \textbf{.683} & .371 & .172 & .419 & .625 & .563 & .385 & .667 & .792 & .515 & .319 & .623 & .789 \\
        BiLSTM-SA & .396 & .182 & \textbf{.474} & \textbf{.694} & .382 & .178 & .438 & .642 & .723 & .621 & .798 & .848 & .695 & .555 & .759 & .856 \\
        \system & .392 & .175 & .478 & .686  & .384 & .181 & \textbf{.440} & .648  & .585 & .421 & .673 & .804 & .562 & .379 & .666 & .809 \\
        \hline
        K-NRM + Phrase & .337 & \textbf{.136} & \textbf{.415} & \textbf{.618} & \textbf{.353} & \textbf{.135} & \textbf{.369} & \textbf{.502} & \textbf{.294} & \textbf{.114} & \textbf{.302} & \textbf{.479} & \textbf{.312} & \textbf{.115} & \textbf{.329} & \textbf{.556} \\
        MV-LSTM + Phrase & \textbf{.388} & \textbf{.180} & \textbf{.454} & \textbf{.679} & \textbf{.360} & \textbf{.157} & \textbf{.410} & \textbf{.621} & \textbf{.699} & \textbf{.611} & \textbf{.772} & \textbf{.837} & \textbf{.657} & \textbf{.500} & \textbf{.759} & \textbf{.831} \\
        CDSSM + Phrase &  .369 & .164 & .431 & .650 & \textbf{.377} & \textbf{.175} & \textbf{.433} & \textbf{.631} & \textbf{.609} & \textbf{.437} & \textbf{.716} & \textbf{.845} & \textbf{.607} & \textbf{.437} & \textbf{.709} & \textbf{.839} \\
        BiLSTM-SA + Phrase & \textbf{.400} & \textbf{.190} & .464 & .676 & \textbf{.404} & \textbf{.186} & \textbf{.455} & \textbf{.653} & \textbf{.727} & \textbf{.632} & \textbf{.791} & \textbf{.879} & \textbf{.706} & \textbf{.582} & \textbf{.798} & \textbf{.875} \\
        \system + Phrase & \textbf{.395} & \textbf{.180} & \textbf{.481} & \textbf{.700} & \textbf{.391} & \textbf{.184} & \textbf{.440} & \textbf{.650} & \textbf{.709} & \textbf{.625} & \textbf{.769} & \textbf{.860} & \textbf{.704} & \textbf{.588} & \textbf{.788} & \textbf{.866} \\
    \end{tabular}
    \caption{Result of click and view predictions when $\mathit{len}=8$}
    \label{tb:click}
    \vspace{-2em}
\end{table*}

